\documentstyle[twoside,fleqn,rgs_workshop,psfig]{article}
\newcommand{\etal}{{\it et al.} }

\newcommand{\rosat}{{\it ROSAT} }
\newcommand{\xmm}{{\it XMM-Newton} }
\newcommand{\chandra}{{\it Chandra} }

\def\lesssim{\mathrel{\hbox{\rlap{\hbox{\lower4pt\hbox{$\sim$}}}\hbox{$<$}}}}
\def\gtrsim{\mathrel{\hbox{\rlap{\hbox{\lower4pt\hbox{$\sim$}}}\hbox{$>$}}}}
                                                                             
\def\mincir{\raise -2.truept\hbox{\rlap{\hbox{$\sim$}}\raise5.truept
\hbox{$<$}\ }}
\def\magcir{\raise -4.truept\hbox{\rlap{\hbox{$\sim$}}\raise5.truept
\hbox{$>$}\ }}    

\begin{document}

\title{RX~J1856.5-3754: BARE QUARK STAR OR NAKED NEUTRON STAR?}

\author{S. Zane
\address{Mullard Space Science Laboratory, UCL, Holmbury St Mary, Dorking, 
Surrey RH5 6NT, UK}
R. Turolla \address{Dept. of Physics, University of Padova, via Marzolo 8, 
35131 Padova, Italy}
J.J. Drake\address{Smithsonian Astrophysical Observatory, MS 3,
60 Garden Street, Cambridge, MA 02138, USA}}

\begin{abstract}

Recent \chandra observations have convincingly shown that the soft X-ray
emission from the isolated neutron star candidate RX J1856.5-3754 is best
represented by a featureless blackbody spectrum, in apparent contrast with
the predictions of current neutron star atmospheric models. Moreover, the
recently measured star distance ($\approx 120-140$~pc) implies a radiation
radius of at most $\sim 5-6$~km, too small for any neutron star equation
of state. Proposed explanations for such a small radius 
include a reduced X-ray emitting region (as a heated polar cap), or the 
presence of a more compact object, as a bare quark/strange star. 
However, both interpretations rely on the presumption that the 
quark star or the cap radiates a pure
blackbody spectrum, and no justification for this assumption has been
presented yet. Here we discuss an alternative possibility. Cool neutron
stars ($T \lesssim 10^6$~K) endowed with a rather high magnetic field ($B
\gtrsim 10^{13}$~G) may suffer a phase transition in the outermost layers.  
As a consequence the neutron star is left bare of the gaseous atmosphere
(``naked''). We computed spectra from naked neutron stars with a surface
Fe composition. Depending on $B$, we found that the emission in the
0.1-2~keV range can be featureless and virtually indistinguishable from a
blackbody. Moreover, owing to the reduced surface emissivity, the star
only radiates $\sim 30-50$\% of the blackbody power and this implies 
that the size of the emitting region is larger than for a perfect 
planckian emitter for the same luminosity. When applied to RX~J1856.5-3754 
our model accounts for the observed X-ray properties and solves the 
paradox of the small radius: we 
predict an apparent star radius of $\sim 10-12$~km, consistent with
equations of state of a neutron star. The optical emission of RX
J1856.5-3754 may be explained by the presence a thin gaseous shell on the
top of the Fe condensate.

\end{abstract}

\maketitle

\section{Introduction}

The family of thermally emitting isolated neutron stars (NSs) includes 
seven peculiar objects serendipitously discovered in \rosat PSPC pointings 
(see e.g. \cite{Treves2000} for a review). 
These sources are characterized by similar properties:
a blackbody-like, soft spectrum with $T_{bb}\sim 100$~eV; low X-ray 
luminosity, $L_X\approx 10^{30}-10^{31} {\rm erg\,s}^{-1}$; low column 
density, $N_H\sim 10^{20} \ {\rm cm}^{-2}$; 
pulsations in the 5-20~s range (detected in four 
sources so far). Due to their proximity and to the fact that they are 
"truly isolated" (i.e. not associated with a supernova remnant), they 
are excellent candidates to probe directly the emission from the star 
surface. 
On the other hand, since they are dim and extremely soft, until recently 
high quality spectra were not available. PSPC
data only provided evidence that a hard tail is absent in all cases and
that a blackbody gives a satisfactory spectral fit. The situation has
recently improved for the two brightest and closest objects, 
RX~J1856.5-3754 and
RX~J0720.4-3125, which have been target of deep observations with \chandra
and \xmm (\cite{Paerels2001,Burwitz2001,Drake2002}). The hope was to 
probe directly the neutron star surface composition, by detecting  
some of the spectral feautures predicted by atmospheric models at 
different chemical composition and/or magnetic field strenght. 
Surprisingly/disappointingly, in both 
sources no spectral features have been detected. In the case of 
RX~J1856.5-3754, the result is particularly robust and striking: 
a very long ($\sim 500$~ks) \chandra pointing has shown that the spectrum 
is better fitted by a simple blackbody than by more sophysticated 
atmospheric models. Also, upper limits on the pulsed fraction of 
RX~J1856.5-3754 are extremely tight, $\mincir 3\%$ 
(\cite{Drake2002}).\footnote{ A few weeks before
this meeting, J. Tr\"umper reported at the 34$^{th}$ COSPAR Scientif
Assembly an even more stringent limit, $\sim 1\%$.  This result is now 
published on \cite{Burwitz2002}.} The sources is known to exhibit 
an ``optical excess'': accurate photometry of the optical 
counterpart with combined Very Large Telescope and Hubble Space Telescope 
data has shown that the UV-optical emission from RX~J1856.5-3754 exceeds 
the Rayleigh-Jeans tail of the X-ray best-fitting blackbody by a factor 
$\sim 6$ (\cite{Walter2002}). The value of the distance ($d \approx 
120-140$ pc; \cite{Kaplan2002,Walter2002}), used in conjunction with 
\chandra data, yields a radiation radius of only $\sim 5$-6 km 
(\cite{Drake2002}), too small for any known NS equation of state. 

The small apparent radius and the blackbody X-ray spectrum led to the
intriguing suggestion that RX~J1856.5-3754 might host a quark star
(\cite{Drake2002,Xu2002}). The motivation is that the radiation radius is
compatible only with equations of state (EOSs) involving strange matter
and that bare quark stars, i.e. those not covered by a layer of ordinary
matter which act as an atmosphere, would {\it presumably} emit a pure 
blackbody spectrum. While a quark star is a conceivable option, present 
observations of RX~J1856.5-3754 do not necessarily demand this solution. 
Other more conventional model fits, based on two blackbodies, can account 
for both the X-ray and optical emission of RX~J1856.5-3754, giving
at the same time acceptable values for the stellar radius
(\cite{Pons2002,Walter2002,Braje2002}). In this scenario X-rays originate
in a relatively large ($\theta \sim 20^\circ$) heated polar cap, while the
rest of the surface is colder ($\sim 30$~eV) and produces the optical
emission. However, since both regions should be covered with an optically 
thick atmosphere, the reason why they should produce a featureless 
spectrum has not been thoroughly understood yet.

In this paper we suggest an alternative possibility. As pointed out by
\cite{Lai1997} (see also \cite{Lai2001}) NSs may be left {\it without an
atmosphere} by the onset of a phase transition that turns the gaseous 
outer
layers into a solid. This happens if the surface temperature drops below a
critical value $T_{crit}$ which in turn depends on the star magnetic
field. The source then appears as a ``naked'' or ``bare'' neutron star: we 
only detect the emission from the solid layers. While
it appears difficult even for the coolest isolated NSs to meet the
requirements for the onset of a phase transition in light element (H, He)
surface layers, this may be the case if the external layers are 
dominated by heavy elements (such as Fe). The
uncertainties on the conditions for Fe condensation are currently quite
large, but, depending on the magnetic field, it is possible that the
surface temperature of RX~J1856.5-3754 falls below the critical value.

If condensation occurs, the emitted spectrum is not necessarily a
blackbody: an overall reduction of the surface emissivity and strong
absorption features are in fact expected when the photon energy becomes
lower than the plasma frequency of the medium. However these absorption
features may or may not appear at soft X-ray energies depending on the
model parameters. In the following we investigate the emission properties
of naked NSs and discuss the relevance of our model in connection with the
NS candidate RX~J1856.5-3754. For all details we refer to
\cite{Turolla2002}.

After this work has been completed we become aware that similar
conclusions have been independently reached by \cite{Burwitz2002} and
have been reported by J. Tr\"umper at the 34$^{th}$ COSPAR Scientif
Assembly. These authors attempted a fit to combined UV-optical and X-ray
data of RX~J1856.5-3754 with a depressed blackbody plus an enhanced
Rayleigh-Jeans tail, by miming phenomenologically the situation here 
described theoretically. Their results for the star radius are close to 
ours.

\section{The Model}
\subsection{Bare Neutron Stars}

Theoretical research on matter in superstrong fields started over
40 years ago and, although many uncertainties still remain,
much progress has been made especially for H and He compositions
(see \cite{Lai2001} for a recent review and references therein). At
$B \gg B_0 = m_ee^3c/\hbar^3 \simeq 2.35 \times
10^9$~G strong magnetic confinement acts on electrons, and atoms
attain a cylindrical shape. It is possible for these
elongated atoms to form molecular chains by covalent bonding along
the field direction. Interactions between the linear chains can
then lead to the formation of three-dimensional condensates. The
critical temperature below which phase separation between
condensed H and vapor occurs is given by \cite{Lai1997}   
\begin{eqnarray}
\label{thyd}
T^H_{crit} & \approx &   0.1 \left [
194.1B_{12}^{0.37} - 4.4 \left ( \ln B_{12} -
6.05 \right )^2  -\hbar \omega_{p,p}  \right.\nonumber \\ 
& -& \left. \frac{\hbar}{2} \left (\omega_{B,p}^2
 +  \omega_{p,p}^2 \right )^{1/2} + \frac{1}{2} \hbar
\omega_{B,p}
\right ]
\,  {\rm eV}\, ,
\end{eqnarray}
where $B_{12}=B/(10^{12}\, {\rm G})$, $\omega_{B,p}$ and $\omega_{p,p}$
are the proton cyclotron and plasma frequencies. For heavier elements
(such as Fe) all estimates are still quite crude. $T_{crit}$ is obtained
by equating the ion density of the condensed phase near zero pressure to
the vapor density, but all available models are approximate near zero
pressure.
The most recent estimate of $T_{crit}$ for Fe phase separation, as given
by \cite{Lai2001}, is $ T_{crit}^{Fe} \approx 27 B_{12}^{2/5} \, {\rm 
eV}$. The density of the condensate (at zero pressure) is $\rho_s
\approx 560 Z^{-3/5} B_{12}^{6/5} \, {\rm g \, cm}^{-3}$
(here $Z$ denotes the atomic number of the constituent element).
This expression should be regarded as accurate to within a factor
of a few while Eqs. (\ref{thyd}) and especially $ T_{crit}^{Fe}$ represent
typical upper limits for the critical temperature.   

\begin{figure}[t] 
\vspace{10pt}
\centerline{\psfig{file=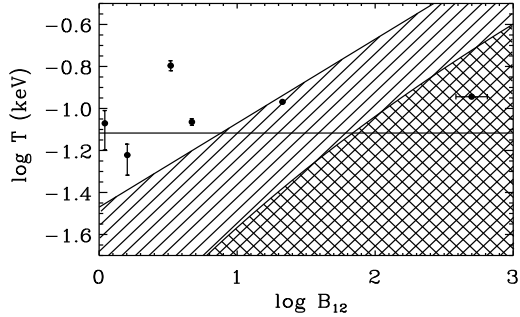,angle=0,width=3.0in,height=1.9in}}
\caption{$T_{crit}$ for H and
Fe as a function of $B$. Condensation is possible in the
shaded region for Fe and in the cross-hatched region for H. Filled circles
mark the position of the sources listed in
Tab.~\ref{tab1}; the horizontal line corresponds to the surface 
temperature of
RX~J1856.5-3754.}
\label{fig1}
\end{figure}

In Fig.~\ref{fig1} we have plotted the critical condensation temperatures
for H and Fe, as a function of $B$, toghether with the 
coolest, thermally emitting NSs for which an estimate of
the magnetic field (as computed from the spin-down formula) is available
(see Tab.~\ref{tab1}). We have also drawn an horizontal line at the 
surface temperature of RX~J1856.5-3754, for which $B$ is
unknown. We used as a color temperature the value
derived from a blackbody fit in the X-rays, $T_{bb}$. The surface
temperature $T_{surf}$ follows by applying the gravitational red-shift
factor.

As we can see from Fig.~\ref{fig1}, most sources have $T_{surf}$ well 
in excess of the H critical temperature. Therefore, if surface layers are
H-dominated, the presence of a gaseous atmosphere is unescapable. The only
exception is RX~J1308.6+2127, for which an exceptionally strong field has
been reported by \cite{Hambaryan2002}. However, this value is probably
preliminary and a re-analysis of the timing properties of this source is
presently under way (Haberl, as reported at the 34$^{th}$ COSPAR meeting).
Similarly, H condensation in RX~J1856.5-3754 requires the star to be a
magnetar ($B \magcir 10^{14}$~G). 

On the other hand, if NSs have not accreted much gas, we might detect
thermal emission directly from the iron surface layers. If this is the
case, for $B \magcir 10^{13}$~G the outermost layers of RX~J1856.5-3754 
might be in the form of hot condensed matter and the usual
radiative transfer computations do not apply.\footnote{The case of
RX~J0720.4-3125 is less certain, because the source falls only marginally
within the region where Fe condensation is possible.} The question of the
nature of the emitted spectrum then arises.

\begin{table}[hbt]
\caption{\centerline{Isolated Neutron Stars Parameters}}
\label{tab1}
\begin{center}
\begin{tabular}{lccl}
\hline
\\
Source & $T_{bb}$ (eV) & $B \, (10^{12}$ G) & Refs.$^a$ \\
\hline
\\      
RX J1856.5-3754 & $61.1\pm 0.3$ &      --       & 
\cite{Burwitz2001,Drake2002} \\   
RX J0720.4-3125 & $86.0\pm 0.6$ & $21.3^{+0.1}_{-0.1} $ & 
\cite{Paerels2001,Zane2002} \\
RX J1308.6+2127 & $90.6\pm 1.6$ & $500^{+150}_{-150}   $ & 
\cite{Hambaryan2002}    
\\
         Vela       & $128.4\pm 7 $ & $3.3   $ & 
\cite{Taylor1993,Pavlov2001}
\\
         Geminga    & $48.3^{+6.1}_{-9.5}$ & $1.5$ & 
\cite{Bignami1996,Halpern1997} 
\\
         PSR 0656+14    & $69.0\pm 2.5$ & $4.7$ &  
\cite{Taylor1993,Marshall2002}      
\\
         PSR 1055-52    & $68.1^{+10.2}_{-17.2}$ & $ 1.1 $ & 
\cite{Taylor1993,Greivendilger1996} 
\\  
\hline
\end{tabular}
\end{center}
\end{table}  

\subsection{The Emitted Spectrum} 

We consider a NS with metallic Fe surface layers and repeat a computation,
first carried out by \cite{Brinkmann1980} and based on the evaluation of
the surface reflectivity, for the parameter range relevant to cold
isolated NSs. At each magnetic co-latitude $\theta$ we first compute the
total reflectivity $\rho_\omega (i,\beta, \theta)$ of the surface element
for incident unpolarized radiation. Here $i$ and $\beta$ are the
co-latitude and the azimuthal angle of the incident wave vector, relative
to the surface normal. The absorption coefficient is then
$\alpha_\omega=1-\rho_\omega$, and Kirchhoff's law is used to derive the
emissivity $j_\omega = \alpha_\omega B_\omega(T)$, where $T$ is the local
temperature and $B_\omega$ is the blackbody function. The monochromatic
flux $dF_\omega$ emitted by the surface element is computed by averaging
over all incident directions, and further integration over the star
surface gives the total flux 
\begin{eqnarray}\label{ftot}
F_\omega& =& \int_0^{\pi}\, dF_\omega =  \int_0^{\pi}B_\omega(T)\sin 
\theta\,
d\theta  \times 
\nonumber \\
& & \int_0^{2\pi}\int_{-\pi/2}^{\pi/2}\alpha_\omega(i,\beta,
\theta)\sin i\, di \, d\beta\, . 
\end{eqnarray} 
In performing the last step we assumed a dipolar field, $B=
B_p[(4-f)\cos^2\theta+f]^{1/2}/2$, where $B_p$ is the polar field strength
and $f\simeq 1.2$ accounts for general-relativistic corrections in a
Schwarzschild space-time. We computed two sets of models either by taking
a constant surface temperature or allowing for a meridional temperature
dependence as given by \cite{Greenstein1983}. We also explored the
parameter space by varing the electron plasma frequency, $\omega_p$,
around $\omega_{p,0} \equiv \omega_p(\rho =\rho_s)$, to account for
possible deviations of the surface density from its zero-pressure value
$\rho_s$ (see e.g. \cite{Lai2001}).
                                            
For $B\magcir 10^{13}$ G, features related to the strong absorption around
the electron cyclotron frequency fall well outside the X-ray range
accessible to the \chandra LETGS and \xmm EPIC-PN, and are of no immediate
interest. However, even above this field, we do not expect the 
spectrum emitted by the solid to be, in general, a blackbody. For $T 
\mincir 100$ eV, the mean energy of the
refracted waves is comparable to (or lower than) the plasma frequency of
the solid. Therefore, we encounter 
modes with very large refractivity (like whistlers in the terrestrial
atmosphere), modes which are frozen in the medium, or modes highly damped 
which just enter the NS crust and disappear within a small penetration 
depth. 
All these effects may  modify the spectrum through the 
production of features and edges, which in turn 
may or may not appear in the X-rays depending on the model parameters.
                                  
\begin{figure}[t] 
\vspace{10pt}
\centerline{\psfig{file=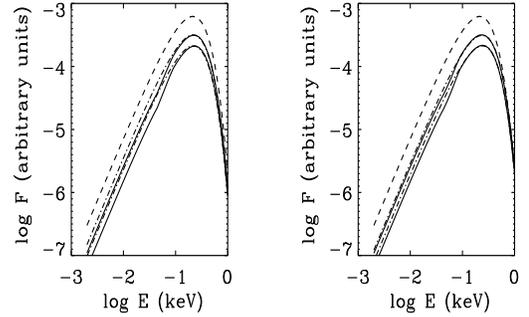,angle=0,width=3.0in,height=1.9in}}
\caption{The spectrum at the star surface for $B_p=3\times
10^{13}$~G, $T_{eff}=75$~eV and two
values of $\omega_p$: $\omega_{p,0}$ (upper solid curve) and
$2.5\omega_{p,0}$ (lower solid curve). Left and right panels refer
to the uniform and meridional temperature distributions, respectively.
The dashed line is the blackbody at $T_{eff}$ and the dash-dotted line the
best-fitting blackbody in the 0.1-2 keV range.
}
\label{fig2}
\end{figure}                                                                             

\begin{figure}[t] 
\vspace{10pt}
\centerline{\psfig{file=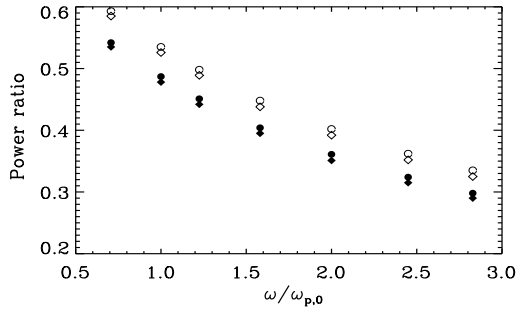,angle=0,width=3.0in,height=1.8in}}
\caption{Ratio of the emitted to the
blackbody power in the 0.1-2 keV band
for different values of the plasma frequency. Circles refer to
$B_p=3\times 10^{13}$ G and diamonds to $B_p=5\times 10^{13}$
G.
Filled and open symbols are for the uniform/meridional variation
temperature distributions respectively.} 
\label{fig3}
\end{figure}                                                                             

In particular we  found that, for $B_p\approx 3-5\times 10^{13}$~G, the 
spectrum in the soft X-rays, (0.1-2~keV) band, shows negligible departures 
(within $\mincir 10\%$) from the best-fitting blackbody and exhibits no 
features whatsoever. We point out that larger deviations are at the
lowest energies, and that the agreement with the planckian is better than 
2\% if we restrict to the interval 0.14-2 keV. 
Spectra are shown in Fig.~\ref{fig2} for $B_p=3\times 10^{13}$~G and
$T_{eff}=75$ eV, which corresponds to 60 eV once the gravitational
red-shift is applied. The constant temperature models have virtually no
hardening, while we get $T_{bb}/T_{eff} \equiv \gamma\simeq 1.14$ 
accounting 
for a meridional temperature variation. An important point is
that, despite being close to planckian {\it in shape}, the spectrum is
substantially depressed with respect to the blackbody at $T_{eff}$, owing
to the reduced emissivity of the surface. The total power radiated by the
star in the 0.1-2 keV band is $\sim$~30-50\% with respect to that of a
perfect blackbody emitter and decreases with increasing $\omega_p$ 
(Fig.~\ref{fig3}).

\section{The Case of RX~J1856.5-3754}

\subsection{The X-ray Emission}

It has been recently proposed by a number of authors 
(\cite{Pons2002,Walter2002,Braje2002}) that the multiwavelength
spectral energy distribution (SED) of RX~J1856.5-3754 may be explained in
terms of a two-temperature surface distribution in a cooling NS.
Although these models appear promising, they are all based on the {\it 
presumption that the emitted spectrum is planckian in shape}. A number 
of mechanisms (magnetic smearing, rotation..) 
have been suggested in order to suppress the spectral 
features predicted by atmospheric models, but {\it no conclusive
evidence}  
has been provided yet that a nearly blackbody, featureless spectrum can be
emitted by an extended atmosphere covering the stellar crust. 
\cite{Braje2002} discuss in detail the role of rotation, showing
that phase-dependent Doppler shifts in a rapidly rotating neutron
star ($P\approx 1$ ms) wash out all features, leaving a nearly
planckian spectrum. Although such a short period can not be
excluded on the basis of present data, the detected periods of
other thermally emitting INSs are in the range $\approx$ 0.1-10 
s, about two orders of magnitude larger. Also, in the 
picture by \cite{Braje2002} the genuine surface temperature should 
correspond to the cooler blackbody at $T\sim 15$ eV. Conventional
cooling curves then imply a star's age of $\approx 10^6$ yr, which is 
hardly compatible with a ms 
period. Furthermore, the energetics of the
bow-shock nebula implies $P = 4.6 \left ( B/10^8~{\rm G}
\right)^{1/2}$~ms. A ms spin period therefore necessary
demands for a very low field star. Such a low field seems hard to
reconcile with the limit on the age derived again from the 
bow shock energetics, $(B/10^{12}~{\rm G})(\tau/10^6~{\rm yr})
\sim 3-4$.

Without a measured period and period derivative, the magnetic field of
RX~J1856.5-3754 is still a mystery. We found that, for the surface layers 
of RX~J1856.5-3754 to be in the form of condensed Fe, its field should be in 
excess of $10^{13}$ G, and more probably at least 3-5$\times 10^{13}$ G
(Fig.~\ref{fig1}). Although rather high, this field strength is well
below the magnetar range and is noticeably shared by another \rosat 
isolated NS: RX~J0720.4-3125 \cite{Zane2002}. When combined with 
the limit
derived from the bow shock energetics, $(B/10^{12}~{\rm G})(\tau/10^6~{\rm
yr}) \sim 3-4$, such field demands for a middle-aged NS ($\tau\approx
10^5$~yr). The spectrum from a bare NS with a dipolar field in this range
is shown in Fig.~\ref{fig2}. In the 0.15-2 keV band the spectrum is
featureless; deviations from the best-fitting blackbody are $\mincir
2\%$, well within the limits of calibration uncertainties of the
\chandra LETGS ($\sim10\%$, as reported by \cite{Braje2002}). 
Our surface emissivity is angle-dependent, and we checked that no 
pulsations are expected to within 3\%, the latest published value 
(\cite{Drake2002}).  

For the radiation radius to be representative of the true star radius the
emission properties of the surface need to be accounted for and, 
at least for thermal emission, we can use the spectral hardening $\gamma$
to quantify spectral deviations from a pure blackbody. Denoting the
fraction of the stellar surface responsible for the observed emission and
the ratio of emitted to blackbody power by $f_A$ and $f_E$, the 
radiation radius is \begin{equation} R_\infty = 4.25\,
f_A^{-1/2}f_E^{-1/2}\gamma^2 d_{100} 
\left(\frac{T_{bb}}{60\, {\rm eV}}\right)^{-2}\, {\rm km,} \,  
\end{equation}
where $d_{100} \equiv d/(100 \, {\rm pc})$. 
The ``angular size'' of RX~J1856.5-3754 reported by \cite{Drake2002} 
is
$R_\infty/(d/100\, {\rm pc}) = 4.12\pm 0.68$ km. If we correct this value
by assuming $d=130$~pc, emission from the entire surface ($f_A =1$) and
uniform temperature ($\gamma=1$), we get $7.7\pm 1.3~{\rm km} \mincir
R_\infty \mincir 9.9\pm 1.6$~km in correspondence to $1\leq
\omega_p/\omega_{p,0}\leq 2.8$ (see Fig.~\ref{fig3}). For the same range 
of plasma frequencies, models with meridional temperature variation give
larger radii, between $9.4\pm 1.5$ km and $12\pm 2$ km. Therefore,
even this simple approach can provide $R_\infty\sim$ 10-12~km, compatible
with (soft) EOSs for $M\sim 1.4 M_\odot$
(\cite{Lattimer2001}).       

\subsection{The UV-Optical Excess}

\begin{figure}[t] 
\vspace{10pt}
\centerline{\psfig{file=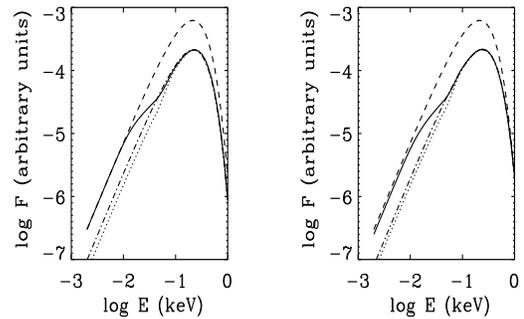,angle=0,width=3.0in,height=1.9in}}
\caption{The multiwavelength SED of a bare neutron star covered by
a thin H layer (full line); the dotted line is the spectrum emitted
by the solid. Details as in Fig.~\ref{fig2}.} 
\label{fig4}
\end{figure}

The UV-optical flux from 
RX~J1856.5-3754 is
enhanced with respect to the extrapolation of the X-ray best-fitting
blackbody. All proposed interpretations for the full SED (from 
optical to X-ray) require two components: either a cold surface plus an 
hotter 
cap or a bare quark star plus a thin crust with flux redistribution. 
Similarly, also our scenario requires the
presence of a second component. A natural possibility is that the
optical flux emitted from the NS surface is reprocessed in a thin, ionized
gas layer on the top of the Fe solid. Hydrogen can be deposited on 
the surface as the result of very slow accretion ($\approx 10^9$~g in $\approx
10^5$~yr). The free-free absorption depth in
the optical ($\sim$ 1-10 eV) is $\approx 10^3$ times larger than in the
X-rays ($\sim 100$ eV), thus in a range of average densities
(around $\approx 10^{-3}\, {\rm g\, cm}^{-3}$) the
gasoues layer is optically thin in the X-rays and thick to optical
photons. 
The scale-height derived from
hydrostatic equilibrium is $\approx 0.1-1$~cm, comparable with the
scale-height at which electron conduction from the crust keeps the gas
almost isothermal at the star surface temperature ($\approx 10^6$~K). 
Preliminary calculation of the multiwavelength SED
are shown in Fig.~\ref{fig4}. Parameters are the same as for the model with
$2.5\omega_{p,0}$ shown in Fig.~\ref{fig3} and the density at the base of 
the layer is $2\times 10^{-3}\, {\rm g\, cm}^{-3}$.
As expected the low-energy SED follows a Rayleigh-Jeans distribution
at the gas (and star) temperature. With increasing energy the optical
depth of the layer drops below unity and therefore X-rays emitted by the
solid surface emerge unchanged from the layer.
However, since the X-ray spectrum emitted by the star
surface is depressed with
respect to the blackbody at the star temperature, the optical
emission appears enhanced. In the model presented here the optical-UV   
excess over the best-fitting X-ray blackbody is a factor $\sim 3$-4,
somewhat lower than the value $\sim 6$ reported by \cite{Walter2002}. 
Larger ratios might be obtained with a slightly larger gas temperature and 
a more thorough investigation is in progress.      

\section{Conclusions}

We have shown that the observed X-ray-to-optical spectra of
RX~J1856.5-3754 can be plausibly explained in terms of the emission from
the surface of a ``naked'' neutron star. The absence of an atmosphere is
due to a phase transition to a solid condensate, which can occur on cool
($< 10^6$~K) neutron stars endowed with strong magnetic fields ($B\magcir
10^{13}$~G) and with metal-dominated outer layers. We have shown that,
when $B\approx 10^{13}$~G, the X-ray spectrum is featureless and planckian
in shape. The observed UV-optical enhanced emission can be explained by
the presence of a gasoues, thin H shell, where the optical flux is
reprocessed.

While we caution that current limitations in our understanding of metallic
condensates in strong magnetic fields renders our estimates of the surface
reflectivity of bare NSs somewhat uncertain, we have
shown that our model for RX~J1856.5-3754 predicts a value for the apparent
radius of $R_\infty\mincir 12$~km. For a canonical NS of mass
$1.4~M_\odot$, such a radius favors soft EOSs.  

\small
\begin{thebibliography}{9}

\bibitem{Treves2000}
Treves A., Turolla, R., Zane, S. \etal, 2000, PASJ, 112, 297

\bibitem{Paerels2001} 
Paerels F. et al., 2001, A\&A, 365, L298 

\bibitem{Zane2002} 
Zane S. et al., 2002, MNRAS, 334, 345 

\bibitem{Burwitz2001}
Burwitz V. et al., 2001, ApJ, 379, L35 

\bibitem{Drake2002} 
Drake J.J. et al., 2002, ApJ, 572, 996 

\bibitem{Burwitz2002}
Burwitz V. et al., 2002, A\&A, in the press

\bibitem{Walter2002} 
Walter F.M. \& J. Lattimer, 2002, ApJ, 576, L145

\bibitem{Kaplan2002}
Kaplan D.L., M.H. van Kerkwijk and J. Anderson, 2002, ApJ, 571, 447

\bibitem{Xu2002} 
Xu R.X., 2002, ApJ, 570, L65

\bibitem{Pons2002}
Pons J.A. et al., 2002, ApJ, 564, 981

\bibitem{Braje2002} 
Braje T.M. \& R.W. Romani, 2002, ApJ in press. (astro-ph/0208069)  

\bibitem{Lai1997}
Lai D. and E.E. Salpeter, 1997, ApJ, 491, 270

\bibitem{Lai2001}
Lai D., 2001, Rev. of Mod. Phys., 73, 629

\bibitem{Turolla2002} 
Turolla R., S. Zane \& J.J. Drake, 2002, ApJ, submitted. 

\bibitem{Hambaryan2002} 
Hambaryan V. et al., 2002, A\&A, 381, 98

\bibitem{Taylor1993}
Taylor J.H., R.N. Manchester and A.G. Lyne, 1993, ApJ Suppl., 88, 529

\bibitem{Pavlov2001} 
Pavlov G.G. et al., 2001, ApJ, 552, L129

\bibitem{Bignami1996}
Bignami G.F. \& P.A. Caraveo, 1996, Ann. Rev. Astron. Astrophys., 34, 331 

\bibitem{Halpern1997} 
Halpern J.P. \& F.Y.-H. Wang, 1997, ApJ, 477, 905

\bibitem{Marshall2002} 
Marshall H.L. \& N.S. Schulz, 2002, ApJ, 574, 377

\bibitem{Greivendilger1996}
Greiveldinger C. et al., 1996, ApJ, 465, L35

\bibitem{Brinkmann1980}
Brinkmann W., 1980, A\&A, 82, 352

\bibitem{Greenstein1983}
Greenstein G., \& G.J. Hartke, 1983, ApJ, 271, 283

\bibitem{Lattimer2001}
Lattimer J.M. \& M. Prakash, 2001, ApJ, 550, 426

\end{thebibliography}

\normalsize
\end{document}